\documentclass[letterpaper, 10 pt, conference]{ieeeconf}
  														                      
\IEEEoverridecommandlockouts                              
                                                          
\usepackage{graphicx}
\begin{document}%
\title{\LARGE \bf Sensors and Navigation Algorithms for Flight Control of
Tethered Kites}%
\author{Michael Erhard$^{1}$ and Hans Strauch$^{2}$%
\thanks{We acknowledge funding from the Federal Ministries {BMWI} and {BMBF},
{LIFE III} of the European Commission, City of Hamburg/BWA and Innovationsstiftung Hamburg.}%
\thanks{$^{1}$M.~Erhard
is with
SkySails GmbH,
Veritaskai 3,
D-21079 Hamburg,
Germany,
e-mail: michael.erhard@skysails.de,
http://www.skysails.de.}%
\thanks{$^{2}$H.~Strauch is a consultant to SkySails.}
}
\maketitle
\begin{abstract}
We present the sensor setup and the basic navigation algorithm used for the flight
control of the SkySails towing kite system.
Starting with brief summaries on system setup and equations of motion of
the tethered kite system, we 
subsequently give an overview of the sensor setup, present the navigation
task and discuss challenges which have to be mastered.
In the second part we introduce in detail the inertial navigation algorithm
which has been used for operational flights for years. 
The functional capability of this algorithm is illustrated by experimental
flight data. 
Finally we suggest a modification of the algorithms as further development
step in order to overcome certain limitations.
\end{abstract}
%
%
\section{Introduction}
The SkySails system is a towing kite system
which allows modern cargo ships to use the wind as source of power in order to
save fuel and therefore to save costs and reduce emissions \cite{WWWSkySails}.
The SkySails company has been founded in 2001 and as main business offers
wind propulsion systems for ships. Starting the development with kites of
6--10\,m$^2$ size the latest generation of products, with a nominal size of
320\,m$^2$, can replace up to 2\,MW of the main engine's propulsion power.
Besides marine applications  there are strongly increasing
activities in using tethered kites 
and rigid wings 
for generating power from high-altitude wind \cite{AWEC2011, Fagiano2012b}. 
Therefore the design of control systems for tethered
kites has become a growing field of experimental \cite{Landsdorp2007a,
Canale2010}
and theoretical
\cite{Ilzhoefer2007, Williams2008c, Furey2007, Baayen2012} 
research efforts.
These control designs demand stable and reliable state estimates
under changing wind conditions 
in order to perform their complex task in a satisfying and robust way.

In this paper we focus on the sensor setup and on the estimation algorithms used for
our flight controller. The most important task is the  estimation of the kite orientation.
In contrast to many applications in the field of inertial navigation we
do not have a sensor which provides absolute attitude information. The task of attitude
estimation, based on acceleration and rotation rate sensors alone, poses a special
challenge for our navigation algorithm.

Many classical inertial navigation algorithms are based on Kalman filtering
\cite{Foxlin1996, Luinge2005, Sabatini2006} and extensions to this
theory \cite{Crassidis2007}.
Our approach is related to complimentary filtering \cite{Mahony2008,
Jensen2011} which will be discussed in the last section.
We would like to emphasize that the aim of this paper is to illustrate
the principle ideas and the challenges to be tackled due to system dynamics and
environmental disturbances specific for our combination of sensors and towing kites.
Thus the presentation also reflects the development 
history and does not intend to be a rigorous, ready-to-use
construction manual for a general type of filter algorithm.

The paper is organized as follows:
We start with short summaries of the overall system setup, the selected coordinates 
and the system dynamics.
We then describe our sensor setup and present the attitude estimation
algorithm in detail. Subsequently we explain the wind referencing 
setup and justify the navigation function by experimental flight data.
We complete the article with a discussion of further
development steps extending the previous algorithms.
\section{System setup and Coordinates}
In the following we will make use of two different coordinate systems for the
navigation and control algorithm, respectively. The control design is based on a
coordinate system with the wind direction as symmetry axis, while the attitude
estimation algorithm is referenced to the gravity (down) direction. The choice
of these different coordinate systems reflect the difference in the two design tasks
(see \cite{Erhard2012a} for further discussion especially of the controller
related aspects).

\begin{figure}
   \centering
   \includegraphics[width=8.6cm]{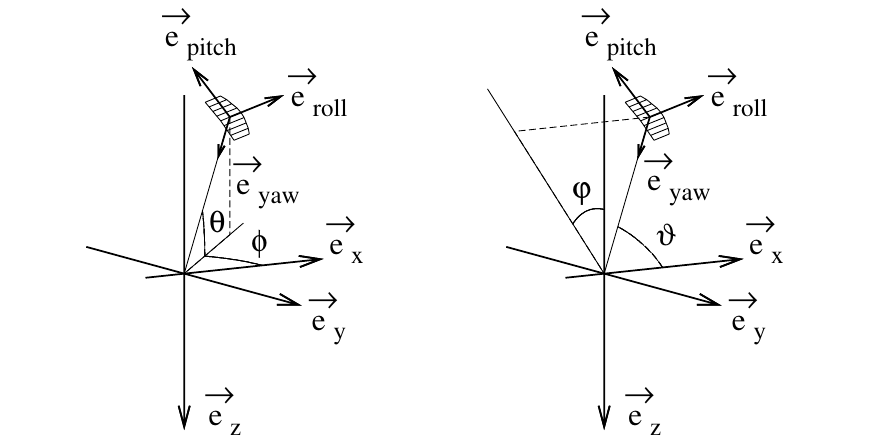}
   \caption{Definition of coordinates for the considered system.
   The right-handed coordinate system is defined by the basis vectors
   $\vec{e}_x, \vec{e}_y, \vec{e}_z$ with $\vec{e}_x$ in wind direction and
   $\vec{e}_z$ pointing downwards with respect to gravity.
   The kite position is parameterized by introducing two different sets
   of spherical coordinates $\left\{\phi,\theta\right\}$ (left figure)
   and $\left\{\varphi,\vartheta\right\}$ (right figure). For a more precise
   definition see (\ref{eq:def_rotation1}) and (\ref{eq:def_rotation2})
   respectively.
   The kite axes are labeled as roll $\vec{e}_{\rm roll}$, pitch $\vec{e}_{\rm
   pitch}$ and yaw $\vec{e}_{\rm yaw}$. This corresponds to the definition usually used
   in aerospace applications with roll axis parallel to forward and yaw axis
   parallel to down directions respectively.
   Note that for the usual flight situation the yaw vector $\vec{e}_{\rm yaw}$
   is defined by the position of the kite assuming it is more or less
   constrained to the origin by a rigid rod.
   Thus orientation of the kite is represented by the single angle
   $\psi_{\rm g}$ and $\psi$ respectively.}
   \label{fig:coordinate_definition}
\end{figure}

The two coordinate systems are introduced in Fig.~\ref{fig:coordinate_definition}.
For a line length $L$ the state of the kite is defined by the set of the three
angles $\left\{\phi, \theta, \psi_{\rm g}\right\}$ and $\left\{\varphi, \vartheta,
\psi\right\}$ respectively. With respect to the basis vectors $\vec{e}_x$,
$\vec{e}_y$, $\vec{e}_z$ the kite position $\vec{x}$
is determined by the set 
$\left\{\phi,\theta\right\}$ and $\left\{\varphi,\vartheta\right\}$ as:
\begin{equation}
    \vec{x} = 
    L \left( \begin{array}{c}
    \cos\phi \cos\theta \\
    \sin\phi \cos\theta \\
    -\sin\theta
    \end{array}\right) = 
    L \left( \begin{array}{c}
    \cos\vartheta \\
    \sin\varphi \sin\vartheta \\
    -\cos{\varphi}\sin\vartheta
    \end{array}\right).\label{eq:kite_position}
\end{equation}
The kite axes are denoted as $\vec{e}_{\rm roll}$ (roll or longitudinal),
$\vec{e}_{\rm pitch}$ (pitch) and $\vec{e}_{\rm yaw}$ (yaw). 

For a description based on rotation matrices one would start with a kite at
position $L\vec{e}_x$, with roll-axis in negative $z$-direction
$\vec{e}_{\rm roll}=-\vec{e}_z$, and then apply the following rotations: 
first $-\psi_{\rm g}$ about $x$, then $\theta$ about $y$ and finally $\phi$ about $z$.
This transformation reads:
\begin{equation}
  R=R_z(\phi)R_y(\theta)R_x(-\psi_{\rm g}) .\label{eq:def_rotation1}
\end{equation}
The quantity $\psi_{\rm g}$ represents the angle between the 
$\vec{e}_{\rm pitch}$-axis and the horizon or the angle between the 
$\vec{e}_{\rm roll}$-axis and the 'upward' $\left(-\vec{e}_z\right)$-direction.

In the second coordinate system $\left\{\varphi, \vartheta,
\psi\right\}$ the rotations are first $-\psi$
about $x$, then $\vartheta$ about $y$ and finally $\varphi$ about $x$.
This transformation reads:
\begin{equation}
  R=R_x(\varphi)R_y(\vartheta)R_x(-\psi). \label{eq:def_rotation2}
\end{equation}
Here one could interpret the angle $\psi$ as orientation of the kite
longitudinal axis with reference to the wind. For a given kite
position $\vec{x}$ (parameterized by $\varphi$ and $\vartheta$) the reference
orientation $\psi=0$ corresponds to the minimum of the scalar product
$\left(\vec{e}_{\rm roll}, \vec{e}_x\right)$ obtained when turning the kite fixated at
this position $\vec{x}$ around its yaw axis $\vec{e}_{\rm yaw}$.
A nonzero value of $\psi$ represents a kite orientation obtained by a rotation of
$\psi$ about the yaw axis $\vec{e}_{\rm yaw}$ starting at this reference orientation.

Next we provide the transformation between the two coordinate systems
$\left\{\phi,\theta,\psi_{\rm g}\right\}\Rightarrow \left\{\varphi,\vartheta,\psi\right\}$:
\begin{eqnarray}
  \varphi &=& \arctan(\sin\phi \cos\theta ,\sin\theta)\label{eq:trafo1_phi} \\
  \vartheta &=& \arccos(\cos\phi \cos\theta)\label{eq:trafo1_theta} \\
  \psi &=& \psi_{\rm g} + \arctan(\sin\phi, \cos\phi \sin\theta).\label{eq:trafo1_psi}
\end{eqnarray}
\section{Dynamics and Equations of Motion}
\begin{figure}
  \centering
  \includegraphics[width=8.6cm]{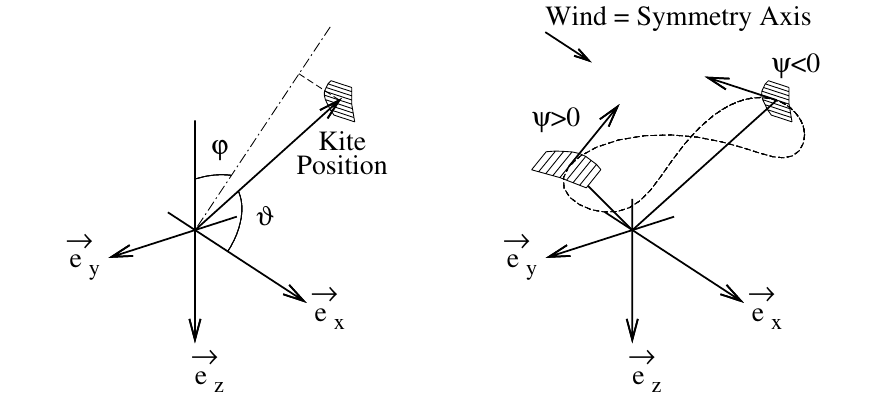}
  \caption{Principle of system dynamics: the kite
  position is given by $\vartheta, \varphi$ (left figure) while the yaw angle
  $\psi$ denotes the orientation with respect to the wind. The quantity
  $\psi$ is crucial for controlling the flight direction of the
  kite, shown for the figure-eight pattern (right figure).}
  \label{fig:kite_dynamics}
\end{figure}

For the description of the  dynamics we refer to Fig.~\ref{fig:kite_dynamics}, illustrating a
dynamically flying kite in a figure-eight pattern.
The position of the kite is parameterized by the previously
introduced coordinates $\varphi, \vartheta$ while the flight direction is
determined by $\psi$. 
Before explaining the dynamics we give a summary of the equations of motion:
\begin{eqnarray}
  \dot{\psi} &=& g\, v_{\rm a}\, \delta + \dot{\varphi}\cos\vartheta\label{eq:psidot}\\
  \dot{\vartheta} &=& \frac{v_{\rm a}}{L} \left(
  \cos\psi - \frac{\tan\vartheta}{E}\right) \label{eq:thetadot}\\
  \dot{\varphi} &=& -\frac{v_{\rm a}}{L \sin\vartheta} \sin\psi.\label{eq:phidot}
\end{eqnarray}
The basic kite response is given mainly by the first term of (\ref{eq:psidot}) 
while $\dot{\varphi}\cos\vartheta$ is a correction due to the motion on the sphere \cite{Erhard2012a}.
A steering deflection $\delta$ results in a turn rate about yaw axis, scaled by
$g$ and air path speed $v_{\rm a}$. The integrated quantity $\psi$ represents
the kite orientation with respect to the wind and thus controls the flight direction
(compare (\ref{eq:thetadot}) and (\ref{eq:phidot})). $E$ denotes the glide ratio of the kite.
For an angle $\psi\!=\!0$ the system stays more or less stationary with
the kite roll axis aligned against the wind direction. In order to obtain the depicted dynamical
'figure eight' pattern one has to command a certain $\psi_{\rm s}\!>\!0$,
resulting in a kite motion into negative $\vec{e}_y$-direction. After some time the commanded
$\psi_{\rm s}$ is changed to a negative value leading to a curve and subsequently to a 
motion into positive $\vec{e}_y$-direction. 

The altitude $\vartheta$ of the flight
trajectory can be controlled by the amplitude of $|\psi_{\rm s}|$. 
We would like to stop at this point and not to proceed into a detailed derivation.
We just conclude that $\psi$ is the crucial quantity for the flight
controller. A detailed justification of the equations of motion and the control
system architecture can be found in \cite{Erhard2012a}.
\section{Sensor Setup}
An overview of the sensor setup is given in Table \ref{tab:sensors}.
\begin{table}
  \centering
  \renewcommand{\arraystretch}{1.3}
  \caption{Overview of sensors aboard the control pod (upper rows) and
  the ship (lower rows).}
  \label{tab:sensors}
  \begin{tabular}{p{2cm}cp{4.5cm}}
  \hline
  Sensor & Quantity & Description\\
  \hline
  Inertial Measurement Unit (IMU) & $\vec{\omega}_{\rm s}$ & Turn rates of the
  control pod below the kite \\
  & $\vec{a}_{\rm s}$ & Accelerations of the control pod below the kite \\
  Impeller Anemometer & $v_{\rm a}$ & Air path speed in $\vec{e}_{\rm
  roll}$-direction
  \\
  Strain Gauge Pod & $F$ & Force measurement towing line  \\
  Barometer & $h$ & Barometric height of the control pod\\
  \hline
  Tow Point & $\phi_{\rm s},\theta_{\rm s}$ & Angles of towing rope with respect to the
  ship (compare Fig.~\ref{fig:coordinate_definition} left).\\
  Ship Anemometer & $v_{\rm w},\phi_{\rm w}$ & Apparent wind speed and direction
  aboard the ship.\\
  Ship IMU &  & Roll and pitch angles of the ship.\\
  Line length & $L$ & Rotary encoder measurement of unwound towing line length\\
  \hline
  \end{tabular}
\end{table}
The data acquisition is performed by  distributed computers  running and
merging the data at a main sample rate of 10\,Hz.

The main sensor is the inertial measurement unit (IMU) in the control
pod, which is situated under the kite including three perpendicular arranged
gyroscopes and accelerometers based on the solid state microelectromechanical
systems (MEMS) technology. For some years we used the Crista OEM sensor
head from Cloudcap Technology Inc.~\cite{Cloudcaptech_CristaOEM} which combines sensors from Analog Devices
Inc.
with an A/D-converter and includes a factory calibrated lookup table for gains
and offsets over the operational temperature range. Later we switched to the {ADIS\,16355}
device from Analog Devices Inc.~\cite{AnalogDevices_ADIS16355} which provides
a complete IMU with temperature calibrated outputs on the digital interface. 

We would like to emphasize at this point that calibration and offset
compensation over the operational temperature range are crucial for the
performance of the navigation algorithm.
In order to improve the acceleration offsets of the ADIS devices we determine
the offset values on a water leveled reference plate before installation of the sensor units.
Both IMUs have a measuring range of $\pm 300$\,deg/s (gyros) and $\pm
100$\,m/s$^2$ (accelerometers). The typical observed bias values ($1\sigma$) for
both IMUs are in accordance with the data sheets and approximately $0.2$\,deg/s
for the gyros and $0.2$\,m/s$^2$ for the accelerometers.

The IMU values are sampled at a rate of 200\,Hz.
20 values are averaged for the turn rates $\vec{\omega}_{\rm s}$ and for the
accelerations $\vec{a}_{\rm s}$ which are then provided to the main 10\,Hz cycle.

The air path speed of the kite in the range of 5--50\,m/s is determined by a frequency
measurement of pulses generated by an impeller anemometer.

For completeness we note that there are also a strain gauge and a
barometer altimeter aboard the control pod. They
are used for redundancy and guidance purposes. However,
these topics will not be elaborated in this paper.

The towpoint is the deflection point of the towing rope aboard the ship. Two
angular sensors determine the mechanically sensed direction of the towing line with respect to
the ship as azimuth and elevation angles. Another IMU device (XSens MTi) is
installed on the ship measuring the roll and pitch angles due to
the wave induced motion of the ship.

The azimuth and elevation angles are merged with the roll and pitch
angles in order to get the azimuth $\phi_{\rm s}$ and elevation $\theta_{\rm s}$
values. This corresponds in good approximation to an inertial
reference and provides reasonable suppression of sea state influences which could
easily cause ship roll amplitudes of $\pm 10$\,deg in operating conditions.
We would like to point out that above transformation is correct only if
the ships' center of angular motion coincides with the tow point, which is usually
not the case. Yet, it is an acceptable approximation as the main
dynamics takes place with respect to the roll axis. 

A further sensor is the anemometer aboard the ship measuring wind speed and
direction of the apparent wind. Although we will argue later that the wind conditions
at flight altitude might be quite different, this sensor is nonetheless an
important reference input for the estimation algorithms especially during the
launch of the system.

The dynamics is influenced by the line length which is
measured by an multiturn rotary encoder attached to the drum of the towing rope
winch.

For our prototype systems we also used
two GPS receivers in the control pod and aboard the ship for calculating the difference
between positions. Although we obtained valuable data for
research and development purposes we discarded the use of  GPS  for operational flights as
the position data suffer from dropouts due to rapid changes of antenna orientation during
dynamic flights.
\section{Yaw Angle Estimator}
\label{sec:yaw_angle_estimator}
In this section we discuss the yaw angle estimator (YAE) which determines
$\psi_{\rm g}$. This quantity can be considered as the attitude angle relative to the horizon.
It is important to emphasize that we have no direct measurement of
this angle, e.g. based on optical or microwave technology. Such a system would be
expensive and difficult to operate at sea. 
Therefore we must estimate the orientation based on the rotation rates
$\vec{\omega}_{\rm s}$ and the acceleration values $\vec{a}_{\rm s}$ alone.

The orientation can be determined by integration of the turn rates but has to be
referenced to the horizon. 
In order to obtain the direction of gravity we carry out the following steps: 
first the measured accelerations are transformed to the integration reference frame
and then the obtained acceleration vector is averaged. This coordinate system
represents, on average, an inertial frame: As we deal with a tethered system 
we expect that due to the averaging all
dynamic accelerations cancel out over time, because otherwise the system 'would fly
away' in contradiction to the constraint by the tether. Thus 
the average value represents an estimate for the gravity vector. However, extra
precautions have to be taken in order to deal with sensor offsets, which may lead
to a slow drift of the coordinate system.

The main idea of the algorithm is illustrated in Fig.~\ref{fig:rotations}: 
\begin{figure}
  \centering
  \includegraphics[width=8.6cm]{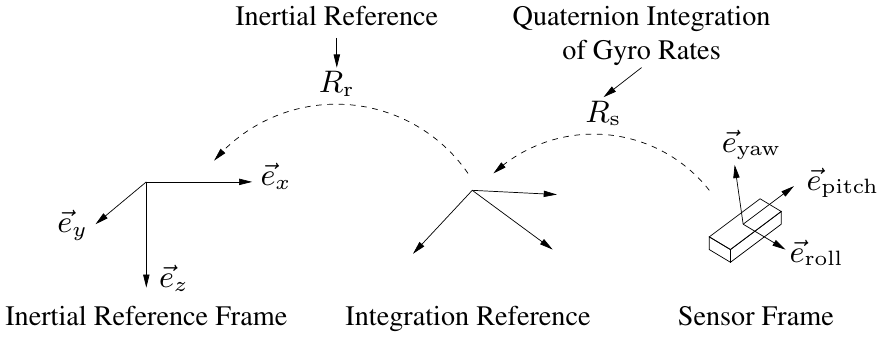}
  \caption{The yaw estimator's state is represented by the two rotations $R_{\rm s}$
  and $R_{\rm r}$. Rotation $R_{\rm s}$ is computed by quaternion integration of
  the three turn rates and depicts the fast dynamics of the flying system. The
  rotation $R_{\rm r}$ references the system to the inertial frame thus 
capturing a potential slow drift due to sensor errors.}
  \label{fig:rotations}
\end{figure}
the turn rates $\vec{\omega}_{\rm s}$ are integrated and then used to build an orientation
matrix $R_{\rm s}$, which represents the orientation of the IMU sensor with
respect to the integration reference. This reference already
features the property of an inertial system but drifts due to the measurement
errors and sensor offsets with respect to the inertial reference frame.
By evaluation of accelerations and extraction  of the apparent
gravity direction a rotation matrix $R_{\rm
r}$ is computed by averaging. Thus $R_{\rm
r}$ relates the, possibly drifting, integration system to the inertial reference frame.
The final navigation angles are
calculated on the basis $R=R_{\rm r}\!\cdot\!R_{\rm s}$ which represents the orientation of
the IMU sensor with respect to the inertial reference system.

\begin{figure}
  \centering
  \includegraphics[width=8.6cm]{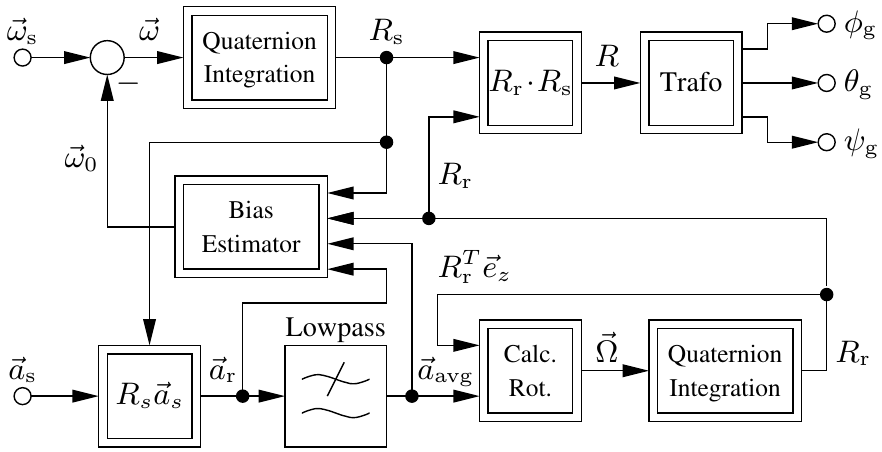}
  \caption{Overview of the yaw angle estimator algorithm which computes the
  orientation estimate $\{\phi_{\rm g}, \theta_{\rm g}, \psi_{\rm g}\}$ using
  turn rates $\vec{\omega}_{\rm s}$ and accelerations $\vec{a}_{\rm s}$.}
  \label{fig:yaw_estimator}
\end{figure}

We now will go into the details of the estimator algorithm. The
respective components and their interactions are shown in
Fig.~\ref{fig:yaw_estimator}.
The first block to be mentioned is the integration of the turn rate
vector $\vec{\omega}\!=\!\vec{\omega}_{\rm s}\!-\!\vec{\omega}_{0}$ to $R_{\rm
s}$.
The vector $\vec{\omega}_{0}$ denotes the sensor bias estimates. Details of the
quaternion propagation can be found 
e.g.~\cite{Kuipers2002a}. The rotation matrix
$R_{\rm s}$ describes the fast dynamics of the systems.

This matrix $R_{s}$ is now used to transform the measured accelerations
$\vec{a}_{\rm s}$ into the integration reference by
\begin{equation}
  \vec{a}_{\rm r} = R_{\rm s} \vec{a}_{\rm s}.\label{eq:a_r}
\end{equation}
This vector is subsequently filtered by a lowpass:
$\vec{a}_{\rm avg} = LP[\vec{a}_{\rm r}]$, 
where $LP[]$ represents three independent filter
operations on the vector components using a second order Butterworth lowpass
with a cutoff frequency of $0.01$\,Hz.  
We would like to emphasize that (\ref{eq:a_r}) and the subsequent
filter realize the central idea introduced in the beginning of this
section: The component wise averaged value $\vec{a}_{\rm avg}$ thus represents an estimate for
the gravity vector.

The next task is to compute the matrix $R_{\rm r}$ such that the normalized
acceleration estimate $\vec{a}_0 = \vec{a}_{\rm avg}/|\vec{a}_{\rm avg}|$ is 
transformed into the $(-\vec{e}_z)$-direction which is the measured acceleration 
signal of the gravitational force pointing into 'down'-direction:
\begin{equation}
  R_{\rm r} \vec{a}_0 = -\vec{e}_z.\label{eq:condition_Rr}
\end{equation}
The straight forward solution of the $R_{\rm r}$ propagation task could be to
determine the drift of this vector between time steps $n\rightarrow (n+1)$ by
$\Delta\vec{a}=\vec{a}_0(n+1)-\vec{a}_0(n)$ and propagate rotation $R_{\rm r}$
accordingly.
Since numerical errors would accumulate a better solution is the
introduction of a feedback loop and basing the propagation upon the deviation
\begin{equation}
   \Delta\vec{a} = \vec{a}_0 -
   R_{\rm r}^T\cdot\left(-\vec{e}_z\right).\label{eq:Delta_e}
\end{equation}
We then have to find a rotation $\vec{\Omega}$ which compensates for
this deviation by turning $\vec{a}_0$ accordingly: 
\begin{equation}
  \Delta\vec{a} = -\vec{\Omega} \times \vec{a}_0.\label{eq:rotation1}
\end{equation}
Inversion of this equation with respect to $\vec{\Omega}$ is not unique. 
If we require $\vec{\Omega} \perp \vec{a}_0$, which is equivalent to the minimal
rotation fulfilling (\ref{eq:rotation1}), some elementary vector algebra yields:
\begin{equation}
  \vec{\Omega} = -\vec{a}_0 \times \Delta\vec{a}. \label{eq:Omega_1}
\end{equation}
This rotation $\vec{\Omega}$ is used for the propagation of $R_{\rm r}$
via quaternion algebra. The navigation angles $\left\{\phi_{\rm g},\theta_{\rm g},\psi_{\rm g}\right\}$
are calculated out of $R=R_{\rm r}\!\cdot\! R_{\rm s}$ by using a similar
sequence of rotations as given in (\ref{eq:def_rotation1}).

For perfect sensors we would have finished the task at this point, 
but our gyroscopes bring
along bias values on the turn rates which would accumulate to angle errors. 
In order to minimize these angle errors, it is crucial to compensate for these
offsets before the quaternion integration (compare
Fig.~\ref{fig:yaw_estimator}). Explaining the estimation of the offset vector
$\vec{\omega}_0$ we
start with the error due to dragging which is defined as:
\begin{equation}
  \Delta\vec{a}_{\rm r} = \vec{a}_{\rm r} - \vec{a}_{\rm avg}.\label{eq:delta_a}  
\end{equation}
Now a rotation $\vec{\Omega}_{\rm r}$ has to be found with the property
\begin{equation}
  \Delta\vec{a}_{\rm r} = \vec{\Omega}_{\rm r} \times \vec{a}_{\rm
  r}.\label{eq:Delta_ar}
\end{equation}
It is worth mentioning that this is a rather intuitive approach as during
curve flights $\left|\vec{a}_{\rm r}\right|\gg\left|\vec{a}_{\rm avg}\right|$.
Therefore the relation to a rotation seems to be weak. But  as
$\vec{\Omega}_{\rm r}$ is strongly filtered the approach holds and worked
in practice. 
An important issue of (\ref{eq:Delta_ar}) is that $\vec{a}_{\rm r}$ is the reference
for $\Delta\vec{a}_{\rm r}$ and not $\vec{a}_{\rm avg}$ in order to get the
smoother behavior. With the same type of reasoning as for (\ref{eq:Omega_1}) we
get
\begin{equation}
  \vec{\Omega}_{\rm r} = \frac{1}{\left| \vec{a}_{\rm r} \right|^2} \left(
  \vec{a}_{\rm r} \times \Delta\vec{a}_{\rm r} \right).
\end{equation}
For estimation of the offset rate $\vec{\omega}_{0}$ the transformed rotation
$R^T_{\rm s} \vec{\Omega}_{\rm r}$  is integrated with a gain of 
$\gamma\!=\!0.003$\,1/s as follows:
\begin{equation}
  \vec{\omega}_{0}(n+1) = \vec{\omega}_{0}(n) + \Delta t \,\gamma\, R^T_{\rm s}
  \vec{\Omega}_{\rm r}.\label{eq:omega_0}
\end{equation}
We are aware that this offset estimation algorithm is more
an intuitive than mathematically rigorous approach. In our description we further omitted
an additional damping term, regarding the rotational degree of freedom with respect
to $\vec{e}_z$, which we have also implemented in order to further robustify the 
filtering scheme. Explaining these technical details cannot be be done within the
limits of this paper.
\section{Wind Referencing}
\label{sec:wind_referencing}
\begin{figure}
  \centering
  \includegraphics[width=8.6cm]{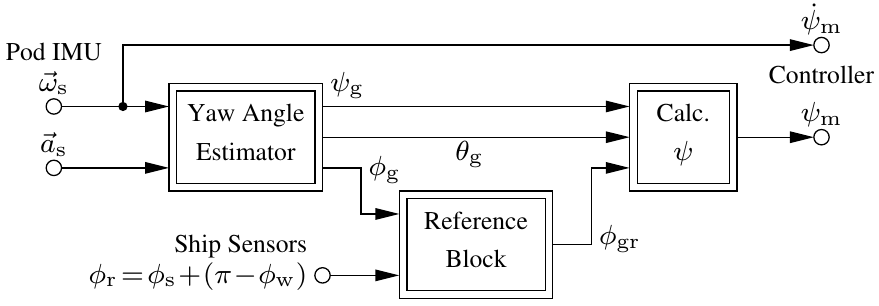}
  \caption{Setup of the navigation algorithm which references the
  $\phi_{\rm g}$ output of the YAE and computes the orientation $\psi_{\rm
  m}$ with respect to the wind for the flight controller.}
  \label{fig:navigation_setup}
\end{figure}

As already introduced, the controller input value $\psi_{\rm m}$ represents the
orientation with respect to the wind. The general arrangement for its
computation is shown in Fig.~\ref{fig:navigation_setup}.
The IMU turn rates $\vec{\omega}_{\rm s}$ and accelerations $\vec{a}_{\rm s}$ are fed into
the yaw angle estimator (YAE) which determines estimates for the angles
$\theta_{\rm g}$ and $\psi_{\rm g}$ with respect to gravity. 
As this gravity reference is a single vector only, the $\phi_{\rm g}$ output
represents the dynamics with respect to 
the vertical $\vec{e}_z$-axis but lacks an
absolute reference and therefore shows an offset drift due to sensor noise.
As a consequence the 'reference block' is introduced in order to reference this
value to $\phi_{\rm r}\!=\!\phi_{\rm s}+(\pi-\phi_{\rm w})$, which is
the horizontal angle of the towing line to the wind direction, calculated from
towpoint angle $\phi_{\rm s}$ and wind direction $\phi_{\rm w}$ aboard the ship.
Without going into details this block performs a complementary filtering
similar to \cite{Mahony2008} with highpass behavior with respect to $\phi_{\rm
g}$ combined with an offset generated by lowpass filtering of $\phi_{\rm r}$.

\begin{figure}
   \centering
   \includegraphics[width=8.6cm]{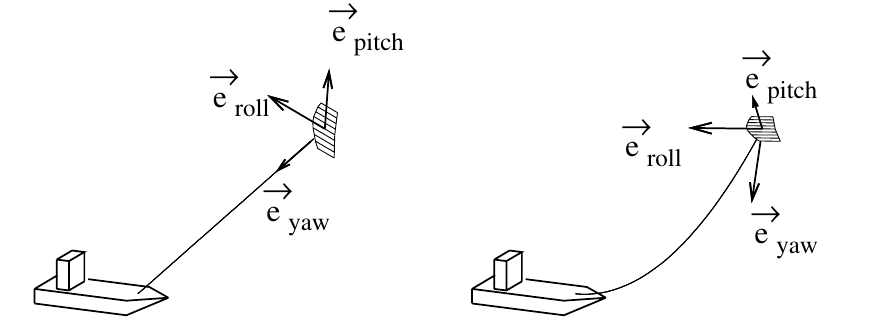}
   \caption{Two different dynamic regimes have to be covered by the navigation
   algorithm.
   In the left figure the regular constrained flight situation is shown. The
   towing line force orients $\vec{e}_{\rm yaw}$ in line direction. For the
   free flight situation in the right figure the orientation of the kite has to
   be considered independently from the position.}
   \label{fig:constrained_freeflight}
\end{figure}

The computation of $\psi_{\rm m}$ is based on $\phi_{\rm gr}$,
$\theta_{\rm g}$ and $\psi_{\rm g}$ using (\ref{eq:trafo1_psi}). The specific
choice of these input quantities is explained with
Fig.~\ref{fig:constrained_freeflight} illustrating the different flight
situations.
In the left figure the regular mode of
operation is shown which is determined by a constrained dynamics. The
$\vec{e}_{\rm yaw}$-vector points into direction of the towing line. Both the
IMU and the towpoint measurement values coincide $\phi_{\rm
gr}\!\approx\!\phi_{\rm s}+(\pi-\phi_{\rm w})$,
$\theta_{\rm g}\!\approx\!\theta_{\rm s}$ and thus can be used likewise in
order to compute $\psi_{\rm m}$ for the control task.

A more challenging situation is the free flight situation shown in the right
part of Fig.~\ref{fig:constrained_freeflight}. Sudden wind fall offs, downwashes
or wave induced motion of the ship may lead to flight situations where the
distance between tow point and kite is shorter than the towing line length.
These situations lead to a completely different dynamics of the flying system
which now more or less behaves like a paraglider. Although in most cases these
situations fortunately last only a few seconds, arrangements must be made that
the input values for the flight controller stay within appropriate
range. 

The computation of $\psi_{\rm m}$ from the data set $\left\{\phi_{\rm
gr},\theta_{\rm g},\psi_{\rm g}\right\}$ 
is based on control pod sensors apart from the referencing of $\phi_{\rm gr}$
to ship sensors. As this referencing takes place on a long-term timescale the 
short-term timescale navigation can be considered as a  local
navigation aboard the control pod which is presumably the best orientation
estimate of the kite with respect to the wind even in extreme flight situations.

We conclude this section by noting that in order to get high-performance
navigation information the determination of the wind direction is a crucial
challenge as wind direction at flight altitude might differ from the wind
direction at sea level up to some tenths of degrees.
Therefore a wind offset estimate is computed which provides an
angle correction value to $\phi_{\rm r}$. Details of this algorithm go beyond
the scope of this paper and will be published elsewhere. 
%
\section{Experimental Validation}
\label{sec:experiments}
\begin{figure} 
  \centering
  \includegraphics[width=8.6cm]{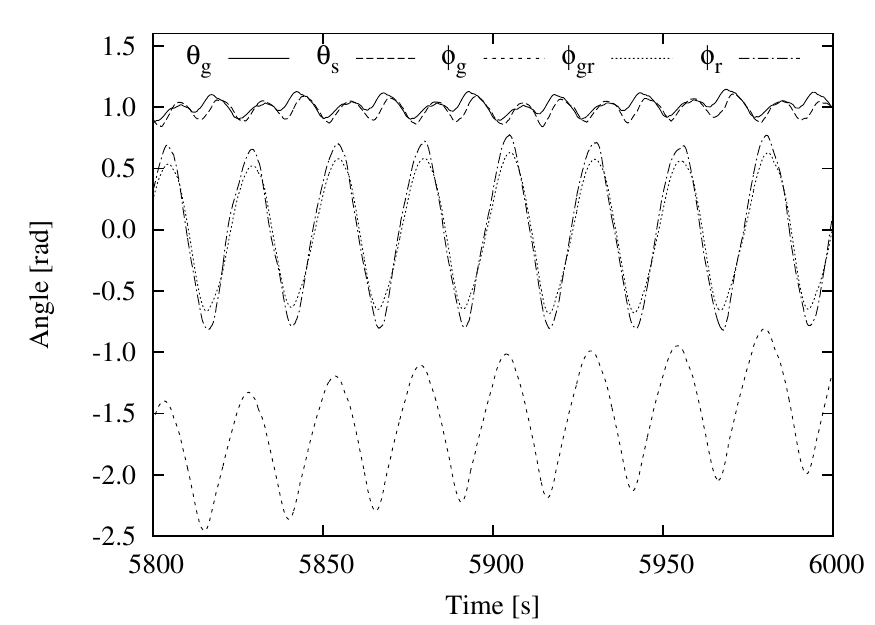}
  \caption{Pattern-eight flight data comparing the YAE output $\phi_{\rm g}$ to
  the reference value $\phi_{\rm r}\!=\!\phi_{\rm s}\!+\!(\pi\!-\!\phi_{\rm
  w})$ computed from ship sensors only. Except for a drifting offset these
  curves are in good agreement. This offset is compensated in the $\phi_{\rm
  gr}$ value. For details see Section \ref{sec:wind_referencing}. Likewise
  the YAE output $\theta_{\rm g}$ coincides with the towpoint value $\theta_{\rm s}$.
  Minor differences are due to additional excitation modes of the towing rope
  and control pod system not included in the simple tether
  assumption.}
  \label{fig:plot02}
\end{figure}

\begin{figure} 
  \centering
  \includegraphics[width=8.6cm]{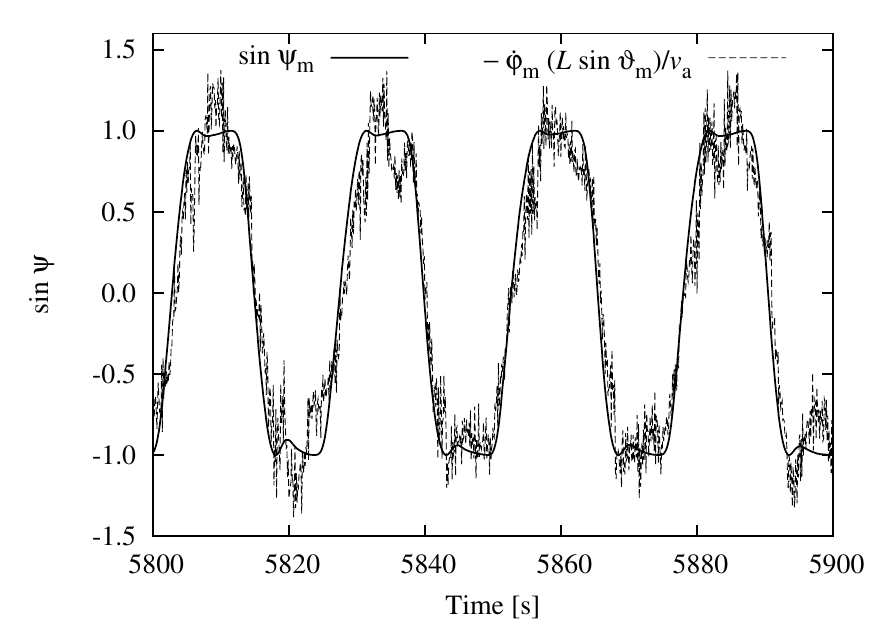}
  \caption{Flight data comparing $\sin \psi_{\rm m}$ to the corresponding
  term of the equation of motion (\ref{eq:phidot}). We give the unfiltered
  signal which is a bit noisy due to the derivation of $\varphi_{\rm m}$ with
  respect to time.}
  \label{fig:plot01}
\end{figure}

In this section we compare the output angles of the YAE to other independent
sensor measurements. As already discussed in Section \ref{sec:wind_referencing}
during tethered dynamic flight 
the
IMU and the towpoint measurement values coincide: 
$\phi_{\rm gr}$ versus $\phi_{\rm s}+(\pi-\phi_{\rm w})$ and
$\theta_{\rm g}$ versus $\theta_{\rm s}$. Corresponding measurement 
curves are shown in Fig.~\ref{fig:plot02}. 
The respective curves are in good agreement as explained in the figure caption.
A verification of the $\psi_{\rm m}$ angle output of the
navigation setup is shown in Fig.~\ref{fig:plot01}. 
The diagram plots the left-hand and right-hand sides
of (\ref{eq:phidot}). The analogy of the curves shows the validity of this
equation and thus the capability of the setup in providing a reasonable $\psi_{\rm m}$ value.
\section{Discussion and Further Development}
\label{sec:discussion}
Although the presented YAE shows convincing results and has been used
successfully for operational flights for years, the approach has 
certain shortcomings in estimating the offset rates $\vec{\omega}_0$.
The fact that the referencing is with respect to the gravity axis only, may lead,
in certain specific flight situations, to
inaccurate offset estimations for sensor axes oriented to the $\vec{e}_z$-axis. 
This issue has also been reported for a Kalman filter approach on IMU data
\cite{Luinge2005}. In order to overcome this situation we suggest a merging of YAE and
wind referencing into one algorithm, thus processing direction of gravity
and wind reference simultaneously. 

\begin{figure}
  \centering
  \includegraphics[width=8.6cm]{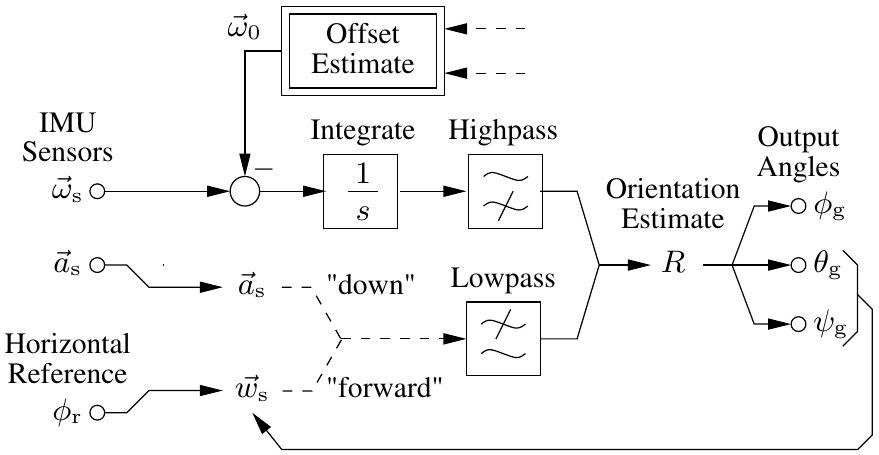}
  \caption{Estimation algorithm combining YAE and wind referencing.
  The orientation input to the lowpass filter is based on the acceleration
  vector $\vec{a}_{\rm s}$ and the wind direction $\vec{w}_{\rm s}$ which is
  computed based on the horizontal reference $\phi_{\rm r}$ and the
  estimator output angles $\theta_{\rm g}$ and $\psi_{\rm g}$. The 
  function blocks are to represent the transfer
  dynamics rather than a concrete algorithm.}
  \label{fig:estimator2}
\end{figure}

An approach of a complementary filter processing accelerometer and
magnetometer information is described in \cite{Mahony2008} which could be
adopted in the following way: instead of the magnetometer
we introduce a fictitious sensor value $\vec{w}_{\rm s}$ emulating a
measurement of the horizontal reference wind direction.
It is obtained by rotating the reference axis $\vec{e}_x$ into the sensor frame
by:
\begin{equation}
  \vec{w}_{\rm s} = {R}(\phi_{\rm r}, \theta_{\rm g}, \psi_{\rm g})^T\,
  \vec{e}_x.
\end{equation}
Note that $\psi_{\rm g}$ and $\theta_{\rm g}$ are filter outputs whereas
$\phi_{\rm r}$ is the horizontal referencing input. 
The principal data flow of such an algorithm is sketched in
Fig.~\ref{fig:estimator2}.

We would like to point out here that we consider the filtering of the
acceleration input as critical: our YAE filter is based on the physics that the
average of the acceleration vectors yields the direction of gravity for a tethered system.
The common approach first calculates a direction out of the acceleration
vector $\vec{a}_{\rm s}$ by normalization and uses 
$\vec{a}_{\rm s}/|\vec{a}_{\rm s}|$ as input value for the filter.
This is also done for filter on the {SO(3)} described in \cite{Mahony2008}.
As we have to deal with accelerations
multiple of the gravity acceleration $|\vec{a}_{\rm s}|\gg g$ during dynamical
flight, it is questionable whether the error introduced by neglecting the
$|\vec{a}_{\rm s}|$-amplitude is acceptable. Modifications and extensions 
to these filters may be necessary.

First validation tests of the combined algorithms with recorded flight data
yielded estimated orientation angles comparable to the output of the operational YAE and
wind referencing setup.
Meanwhile the estimated sensor biases seem to be more realistic than
those of the YAE. However a critical sensitivity of the bias estimator
on filter frequencies and flight situations has been observed, which necessitate
further research and development efforts on these approaches.

Conclusively we would like to note that gains and filter frequencies of the
presented algorithms were determined mainly by trial and error on recorded
flight data sets.
On the other hand a thorough Kalman filter design would be desirable and may 
result in some improvements of the estimator performance. 
However, such a Kalman theory based filter design for a tethered kite system turns out to be non-trivial
as measurements are less corrupted by sensor noise than by  
disturbances (mainly wind), which can not be easily modeled within a statistically based framework.
\bibliographystyle{IEEEtran}
\bibliography{IEEEabrv,mybibfile}
\end{document}